\documentstyle{elsart}

\newcommand{\bea}{\begin{eqnarray}}
\newcommand{\eea}{\end{eqnarray}}
\newcommand{\beq}{\begin{equation}}
\newcommand{\eeq}{\end{equation}}
\newcommand{\vect}[1]{\mbox{\boldmath${#1}$}}
\newcommand{\per}{\delta\!}
\newcommand{\ssA}{\scriptscriptstyle A}

\begin{document}
\begin{frontmatter}
\title{The spectrum of cosmological perturbations \\ produced
  by a multi-component inflaton \\ to second order in the slow-roll
  approximation}
\author[TTN]{Takahiro T. Nakamura\thanksref{ttnmail}} \& 
\author[EDS]{Ewan D. Stewart\thanksref{edsmail}}
\address[TTN]{Department of Physics, University of Tokyo,
  Bunkyo-ku, Tokyo 113, Japan}
\address[EDS]{Research Center for the Early Universe, University
  of Tokyo, Bunkyo-ku, Tokyo 113, Japan}
\thanks[ttnmail]{nakamura@utaphp7.phys.s.u-tokyo.ac.jp}
\thanks[edsmail]{eds@utaphp7.phys.s.u-tokyo.ac.jp}
\begin{abstract}
\noindent
We derive general analytic formulae for the power spectrum and
spectral index of the curvature perturbation produced during inflation
driven by a multi-component inflaton field, up to the second order in
the slow-roll approximation.  We do not assume any specific properties
of the potential or the metric on the scalar field space, except for
the slow-roll condition, Einstein gravity, and the absence of any
permanent isocurvature modes.
\end{abstract}


\end{frontmatter}

\unitlength=1mm
\begin{picture}(100,0)(0,0)
\put(-10,135){\shortstack[l]{\sl
UTAP: University of Tokyo, Theoretical Astrophysics \\ \sl
RESCEU: Research Center for the Early Universe}
}
\put(135,130){\shortstack[l]{UTAP-228/96\\ RESCEU-9/96
\\ astro-ph/9604103}}
\end{picture}
\section{Introduction}

\noindent
Inflation of the early universe \cite{LKT} magnifies microscopic
quantum fluctuations in the inflaton field $\phi$ into macroscopic
classical perturbations in space-time and matter. The latter are
supposed to be the seeds that grow to become the rich structures, such
as galaxies or clusters of galaxies, that are observed today. Thus the
power spectrum $P$ and spectral index $n$ predicted by a model of
inflation can be tested from observation of the large-scale
structures, and it is therefore important to calculate them as
accurately as possible.

Standard calculations \cite{BST} of $P$ and $n$ work to lowest order
in the slow-roll approximation, and assume that $\phi$ has only one
dynamical degree of freedom. However, the latter assumption has no
theoretical or observational justification. Previously, Stewart \&
Lyth \cite{SL} computed $P$ and $n$ up to the second order in the
slow-roll approximation, in the single component inflaton case.
Sasaki \& Stewart \cite{SS} derived general formulae for $P$ and $n$
in the multi-component inflaton case, but only up to the first order
in the slow-roll approximation. In this paper, we calculate $P$ and
$n$ for a multi-component inflaton up to the second order in the
slow-roll approximation. The method of our calculation also refines
the earlier one in Ref.\cite{SS}. We use the units $c=\hbar=8\pi G=1$.

\section{Homogeneous background and slow-roll approximation}

\noindent
Let $\phi^a$ be the multiple scalar fields that slowly roll on the
potential $V(\phi)$ during inflation.  We start from the action of the
form:
\beq\label{1}
S=\int d^4x\sqrt{-g}
\left[\frac12 h_{ab} g^{\mu\nu} \partial_{\mu}\phi^a
\partial_{\nu}\phi^b  - V(\phi) - \frac12 R \right]
\eeq
where $g_{\mu\nu}$ and $R$ are the metric and curvature scalar in
space-time, and $h_{ab}$ is the metric on the $\phi$-space (which may
be curved). In an exactly homogeneous universe, we have
$\partial_i\phi^a=0$ ($i=1,2,3$) and
\beq\label{1.1}
ds^2:=g_{\mu\nu}dx^{\mu}dx^{\nu}
 = dt^2 - a^2(t)\delta_{ij}dx^i dx^j
\eeq
(we assume that the background universe is spatially flat). Then, from
Eqs.\ref{1} and \ref{1.1}, $\dot\phi^a := \partial_0\phi^a(t)$ obeys
the equation
\beq\label{2}
\ddot\phi^a + 3H \dot\phi^a + V^{,a} = 0
\eeq
where $\ddot\phi^a := (D/dt)\dot\phi^a := \dot\phi^b\nabla_{\!
  b}\dot\phi^a$, $\nabla_{\!\! a}$ is the covariant derivative
operator associated with $h_{ab}$, and $H:=\dot a/a$ is the Hubble
parameter. Raising and lowering of indices $abc\cdots$ are always done
by $h_{ab}$. Also we have from the Einstein equation
\beq
3H^2 = \frac12 \dot\phi^a \dot\phi_a + V\,, \label{3}
\eeq
\beq
\dot H = -\frac12 \dot\phi^a \dot\phi_a\,. \label{4}
\eeq
We assume that the potential $V(\phi)$ has a sufficiently gentle
slope:
\beq
V_{,a}V^{,a} \ll V^2, \quad 
(V_{;ab}V^{;ab})^{1/2} \ll V\,, \mbox{\ etc.}
\eeq
(we list all the assumptions we need in this paper more rigorously in
Appendix \ref{sec:sr}). Then, from Eqs.\ref{2} and \ref{3},
$\phi^a(t)$ soon approaches a slowly rolling state given by
\beq
\frac{\dot\phi^a}H \simeq -\frac{V^{,a}}{3H^2}
\simeq -\frac{V^{,a}}{V}\,.
\eeq
Let us define
\beq
\alpha := -\frac{\dot H}{H^2}\,, \quad
\beta := \frac{\ddot\phi^a\dot\phi_a}{H \dot\phi^a\dot\phi_a}\,.
\eeq
These are small quantities ($\ll 1$) of the same order in the
slow-roll approximation. They are slowly varying and their time
derivatives (divided by $H$) are smaller quantities of the next order.
For example
\beq\label{10}
H^{-1}\dot\alpha = 2\alpha(\alpha + \beta)
\eeq
is a second order quantity. We use the notation $\simeq$ when the
equality is valid only up to the lowest order in the slow-roll
approximation.

\section{Perturbation}

\noindent
A scalar perturbation in the space-time metric is most generally
written as \cite{KS}
\beq\label{11}
ds^2= (1+2A)dt^2 - 2a(\partial_i B) dtdx^i
- a^2[(1+2{\cal R})\delta_{ij}
+ 2\partial_i\partial_j E]dx^idx^j.
\eeq
Here ${\cal R}$ is interpreted as the intrinsic-curvature perturbation
in the constant time hypersurface. Let $\per\phi^a$ be the
perturbation in the scalar fields around $\phi^a(t)$. In appendix
\ref{app:a}, we derived from Eqs.\ref{1} and \ref{11} the equation of
motion of $\per\phi^a$ on flat hypersurfaces:
\beq\label{9}
\frac{D^2}{dt^2}\per\phi^a + 3H \frac{D}{dt}\per\phi^a 
+ R^a{}_{cbd} \dot\phi^c \dot\phi^d \per\phi^b
+ \left(\frac{k}a\right)^2\per\phi^a + \per\phi_b V^{;ab} 
= \frac{\per\phi_b}{a^3} 
\frac{D}{dt}\left[ \frac{a^3}H\dot\phi^a\dot\phi^b \right]
\eeq
where $R^a{}_{bcd}$ is the Riemannian curvature tensor in the
$\phi$-space. We work in $\vect k$-space throughout and simply use
$\per\phi^a$ as the Fourier transform of the perturbation. The
conformal time $\eta$ is defined by
\beq\label{14}
\eta := \int \frac{dt}a = -\frac1{aH} 
+ \int \frac{\alpha da}{a^2 H}\,.
\eeq
Since $\alpha$ is slowly varying (see Eq.\ref{10}), we take $\alpha$
out of the integral and obtain
\beq
\eta = -(1+\alpha)/(aH)\,.
\eeq
Defining $u^a : = a\per\phi^a$ and working to the first order in
$\alpha$, $\beta$, etc., Eq.\ref{9} is rewritten as
\beq\label{13}
\frac{D^2}{d\eta^2}u^a + k^2u^a = \frac1{\eta^2} (2u^a +
3\epsilon^a_b u^b) 
\eeq
where we regard
\beq
\epsilon_{ab} :=
\alpha h_{ab} + \left( h_{ac}h_{bd} 
- \frac13 R_{acbd} \right) \frac{\dot\phi^c\dot\phi^d}{H^2}  
- \frac{V_{;ab}}{3H^2}
\eeq
as a first order quantity. In order to solve the differential equation
\ref{13}, we introduce the orthonormal basis $e^a_{\ssA}$ ($A$ runs
over the number of scalar field components) parallel-transported along
the unperturbed trajectory $\phi^a(t)$:
\beq\label{16}
\frac{D}{d\eta} e^a_{\ssA} = 0
\eeq
so that the symmetric tensor $\epsilon_{ab}$ is diagonalized along
$\phi^a(t)$ as
\beq\label{17}
\epsilon^{ab} \simeq \sum_{\ssA} \epsilon_{\ssA}
(e^a_{\ssA} \otimes e^b_{\ssA})\,.
\eeq
This diagonalization is justified as follows. At some point on
$\phi^a(t)$, $\epsilon_{ab}$ can be diagonalized exactly as in
Eq.\ref{17} with the eigen-vectors $e^a_{\ssA}$. As one moves along
$\phi^a(t)$ with $e^a_{\ssA}$ parallel-transported, $e^a_{\ssA}$ will
not remain the eigen-vectors and off-diagonal components may appear in
Eq.\ref{17}. However, since we are assuming (see Eq.\ref{ass1}) that
$\epsilon_{ab}$ is covariantly changing slowly along $\phi^a(t)$, the
off-diagonal components are second order quantities. Therefore,
Eq.\ref{17} is valid up to the lowest order in the slow-roll
approximation. In short, we treated $\epsilon_{ab}$ as a constant in
Eq.\ref{17}, just as we treated $\alpha$ as a constant in Eq.\ref{14}.
From Eqs.\ref{16} and \ref{17}, the $A$-component of Eq.\ref{13} is
written as
\beq\label{18}
\frac{d^2u_{\ssA}}{d\eta^2} + k^2 u_{\ssA} 
= \frac1{\eta^2} ( 2 + 3\epsilon_{\ssA}) u_{\ssA}
\eeq
where $u_{\ssA}:= u_a e^a_{\ssA}$. First let us consider microscopic
fluctuations, the physical wavelength of which is well-inside the
horizon ($-k\eta\to\infty$). When $-k\eta\to\infty$, the right hand
side (RHS) of Eq.\ref{18} is negligible compared with the $k^2$ term,
and thus the $u_{\ssA}$ behave like (real) massless Klein-Gordon
fields:
\beq\label{23}
u_{\ssA}(\vect k) = \frac1{\sqrt{2k}} \left[a_{\ssA}(\vect k)
e^{-ik\eta} + a_{\ssA}^{\dagger}(-\vect k) e^{ik\eta} \right]
\eeq
where $a_{\ssA}^{\dagger}$ and $a_{\ssA}$ are the creation and
annihilation operators of an $A$-particle:
\beq
[a_{\ssA}(\vect k), a_{\scriptscriptstyle B}^{\dagger}(\vect k')] =
\delta_{\scriptscriptstyle AB}\delta^3(\vect k -\vect k'), 
\quad a_{\ssA}|0\rangle = 0.
\eeq
As $-k\eta$ approaches unity, the RHS of Eq.\ref{18} becomes
comparable to the $k^2$ term, and the solution is written in terms of
the Hankel functions as
\beq\label{19}
u_{\ssA}(\vect k)= (-k\eta)^{1/2}
\left[C_{\ssA}(\vect k) H^{\scriptscriptstyle (1)}_{\nu_A}(-k\eta) 
+ C_{\ssA}^{\dagger}(-\vect k)
H^{\scriptscriptstyle(2)}_{\nu_A}(-k\eta) 
\right]
\eeq
where
\beq
\nu_{\ssA} := \frac32 + \epsilon_{\ssA}\,.
\eeq
Using the asymptotic behavior of the Hankel functions at infinity, the
integral constants $C_{\ssA}$ are determined from Eq.\ref{23} as
\beq
C_{\ssA}(\vect k) = \sqrt{\frac{\pi}2}
\exp\left[\frac{i\pi}4 (2\nu_{\ssA} + 1)\right]
\frac{a_{\ssA}(\vect k)}{\sqrt{2k}}\,.
\eeq
Next we go well-outside the horizon, i.e., $-k\eta\to0$ [but
$-\ln(-k\eta)$ is not too large]. Using $H^{\scriptscriptstyle
  (1,2)}_{\nu}(x)\to \pm {\mit\Gamma(\nu)}(2/x)^{\nu}/(i\pi)$ as
$x\to0$, and expanding up to the first order in $\epsilon_{\ssA}$, one
finds
\bea
u_{\ssA}(\vect k) &\to& \frac{i}{\sqrt{2k}} \left(\frac{-1}{k\eta}\right)
\frac{{\mit\Gamma}(\nu_{\ssA})}{{\mit\Gamma}(3/2)}
\left(\frac{-2}{k\eta}\right)^{\epsilon_A} b_{\ssA}(\vect k) \\
&=& \frac{i}{\sqrt{2k}}\left(\frac{-1}{k\eta} \right)
\left\{ 1 + [c -\ln(-k\eta) ]\epsilon_{\ssA} \right\} b_{\ssA}(\vect k)
\label{26}
\eea
where 
\beq
b_{\ssA}(\vect k) := e^{i\pi\epsilon_A/2} a_{\ssA}(\vect k) -
e^{-i\pi\epsilon_A/2} a_{\ssA}^{\dagger}(-\vect k)\,,
\eeq
and $c := 2-\ln2-\gamma = 0.7296\cdots$ with the Euler number
$\gamma$. It is clear from Eq.\ref{26} that the perturbations become
completely classical as we go outside the horizon, because $a_{\ssA}$
and $a^{\dagger}_{\ssA}$ appear only in the combination of $b_{\ssA}$
and hence $[u_{\ssA}, \dot u_{\ssA}]=0$ follows. Going back to general
coordinates, we obtain
\beq\label{27}
\per\phi^a = \frac{i H}{\sqrt{2 k^3}} 
\left\{ (1 - \alpha)h^a_b 
+ \left[c + \ln\left(\frac{aH}k\right)\right] \epsilon^a_b \right\}b^b
\eeq
with $b^a := \sum_{\ssA}b_{\ssA}e_{\ssA}^a$. For later use, we note
that
\beq\label{32}
\langle b_a(\vect k) b_b^{\dagger}(\vect k') \rangle 
= h_{ab}\delta^3(\vect k-\vect k')
\eeq
where $\langle\cdots\rangle$ reads the vacuum expectation value.

\section{Power spectrum and spectral index of $\cal R_{\rm c}$}

\noindent
Sasaki \& Stewart \cite{SS} showed that the curvature perturbation on
a comoving hypersurface $\cal R_{\rm c}$ (see Eq.\ref{11}) during the
radiation-dominated phase (after complete reheating) is related to
$\per\phi^a$ as
\beq\label{28}
{\cal R}_{\rm c} = N_{\!,a}\per\phi^a\,.
\eeq
The RHS is to be evaluated at some time (say $t_1$) during inflation
soon after the scale of the perturbation goes well-outside the horizon
(but does not depend on the exact value of $t_1$, as shown below). It
is also assumed in Eq.\ref{28} that the space-time is foliated on a
flat hypersurface at $t_1$, in accord with Eq.\ref{9}. Here
\beq\label{29}
N(\phi) := \int_{t_1(\phi)}^{t_2} H dt
\eeq
is the number of $e$-folds in the background universe, and $t_2$ is
the time corresponding to some fixed energy density during the
radiation-dominated phase.  In general $N$ can depend on both
$\phi^a(t_1)$ and $\dot\phi^a(t_1)$. However, as we are assuming that
slow-roll has been achieved, the $\dot\phi$-dependence should be
eliminated using the slow-roll trajectory which is given in
Eq.\ref{phidot} up to second order.
The power spectrum $P(k)$ of $\cal R_{\rm c}$ is defined by
\beq
\langle {\cal R_{\rm c}(\vect k) R^{\dagger}_{\rm c}(\vect k')} \rangle 
= 2\pi^2 k^{-3} P(k) \delta^3(\vect k -\vect k')\,.
\eeq
From Eqs.\ref{27}, \ref{32} and \ref{28}, one finds
\beq\label{33}
P(k) = N^{\!,c}N_{\!,c} \left(\frac{H}{2\pi}\right)^2
\left\{ 1 - 2\alpha + 2\left[c +
\ln\left(\frac{aH}k\right) \right]\epsilon_{ab}M^{ab} \right\}
\eeq
where $M_{ab}:=N_{\!,a}N_{\!,b}/N^{\!,c}N_{\!,c}$. Thus the spectral
index is
\beq
n := 1 + \frac{d\ln P}{d\ln k} = 1 - 2\epsilon_{ab}M^{ab}
\eeq
which is the identical result with Ref.\cite{SS}. 
In order to calculate $n$ up to the second order, we rewrite
Eq.\ref{33} to show more explicitly that $P(k)$ does not depend on
time. We expand the prefactor of Eq.\ref{33} as
\beq\label{37}
N^{\!,c}N_{\!,c}H^2 = (N^{\!,c}N_{\!,c}H^2)_{aH=k} 
\left[1 + \left. \frac{d\ln(N^{\!,c}N_{\!,c}H^2)}{d\ln
  a}\right|_{aH=k} \ln\left(\frac{aH}k\right)\right]\,.
\eeq
Substituting Eq.\ref{37} into \ref{33}, and using Eq.\ref{b12}, we see
that the $\ln(aH/k)$ terms cancel and obtain
\beq\label{38}
P(k) = \left.N^{\!,c}N_{\!,c} \left(\frac{H}{2\pi}\right)^2
( 1 - 2\alpha + 2c\epsilon_{ab}M^{ab} ) \right|_{aH=k}\,.
\eeq
In this expression for $P(k)$, the $k$-dependence of the LHS is such
that the RHS (which is a function only of time) is evaluated at the
horizon-crossing time $aH=k$. Thus $P(k)$ does not depend on time, as
noted above. To avoid any confusion, let us define $Q(a)$ to be the
RHS of Eq.\ref{38} so that
\beq
P(k) = Q(a)|_{aH=k}\,.
\eeq
Then, using Eqs.\ref{10}, \ref{b11} and \ref{b14}, $n$ is calculated up
to second order as
\bea\label{39}
n &=& 1 + \frac{d\ln Q}{d\ln aH} = 
1 + (1+\alpha)\frac{d\ln Q}{d\ln a} \\
&=&
1 - 2\alpha + 2\lambda_{ab}M^{ab} -2(3-2c)\alpha^2 
-4(1-c)\alpha\beta
+ \frac83\alpha\lambda_{ab}M^{ab}
+ 4c(\lambda_{ab} M^{ab})^2 
\nonumber \\ &&
- \frac23(6c-1) M_{ab}\lambda_a^c\lambda_{bc}    
-\frac43\frac{N^{\!,a}}{N^{\!,c}N_{\!,c}}\frac{\dot\phi^b}H
\frac{V_{;ab}}{3H^2} 
- \frac{2\alpha}{N^{\!,a}N_{\!,a}}
- \frac23(3c+1) \frac{M^{ab}}H \frac{D}{dt}\lambda_{ab}
\label{44}
\eea
where
\bea\label{lambda}
\lambda_{ab} &:=& \left( \frac13 R_{acbd} -h_{ac}h_{bd}
\right) \frac{\dot\phi^c\dot\phi^d}{H^2} + \frac{V_{;ab}}{3H^2} \\ 
&=& \alpha h_{ab} - \epsilon_{ab}\,.
\eea
[Note that $1/(N^{\!,a}N_{\!,a})$ is a first order quantity.] The
$k$-dependence of $n$ is understood in the same way as in Eq.\ref{38}.
Rewriting Eq.\ref{44} in terms of $V$, we find
\bea
&& \hspace{-20pt} n =  
 1 - U^{,a}U_{,a} + 2M^{ab}W_{ab} - \frac12(U^{,a}U_{,a})^2 
- \frac23(3c-2)U^{;ab}U_{\!,a}U_{\!,b} 
+ M^{ab}U_{\!;ab}U^{\!,c}U_{\!,c}
\nonumber \\ &&
+ 4c (M^{ab}W_{ab})^2 - \frac23(6c-1)M^{ab}W_{ac}W^c_b 
+ \frac23 (3c+1) M^{ab} [U_{\!;abc}U^{\!,c}
+ \frac13 R_{acbd;e} U^{\!,c}U^{\!,d}U^{\!,e}]
\nonumber \\ &&
+ \frac49 M^{ab}R_{acbd} U^{\!,c}U_{\!,e}[U^{\!,e}U^{\!,d} 
+ (3c+2) U^{;ed}]
\label{45}
\eea
where
\beq
U := \ln V \,,
\eeq
\beq
W_{ab}:= U_{;ab} + \frac13 R_{acbd}U^{,c}U^{,d} \simeq \lambda_{ab}\,.
\eeq
($U_{\!,a}U_{\!,b}$ and $U_{\!;ab}$ are first order quantities.)

\section{Summary}

\noindent
We have derived general analytic formulae for the power spectrum $P$
(Eq.\ref{38}) and spectral index $n$ (Eq.\ref{44} or \ref{45}) of the
curvature perturbation $\cal R_{\rm c}$ produced during inflation
driven by a multi-component inflaton field, up to the second order in
the slow-roll approximation.
Once one specifies a model of inflation and calculates the number of
$e$-folds $N(\phi)$ (Eq.\ref{29}) in the background universe, then the
substitution of them into our general formulae immediately yields the
power spectrum and its index with accuracy.  We anticipate that, to
lowest order, $N_{\!,a}$ can be calculated from the inflationary phase
only. However, to the next order, as considered in this paper, the
contribution to $N_{\!,a}$ from the reheating and radiation-dominated
phases should be significant.

The magnitude of the first order terms is of order $N_{\rm end}^{-1}$
in many inflation models ($N_{\rm end}$ is the number of $e$-folds
from the horizon-crossing time to the end of inflation), and if
thermal inflation \cite{LS} occurs after ordinary inflation, we have
$N_{\rm end}\sim 30$--40. In this case, the correction terms in
Eq.\ref{38} should be observable, while those in Eqs.\ref{44} and
\ref{45} may be marginally observable. Thus, our formulae may be
useful in testing models of inflation when $P$ and $n$ are observed
accurately by presently-planned experiments such as the {\em Microwave
  Anisotropy Probe}. At the same time, the theories and models of
inflation need to progress to make more precise predictions.

\bigskip\noindent
We would like to thank N.~Sugiyama for helpful discussions, and
D.~H.~Lyth and Y.~Suto for useful comments. TTN and EDS are supported by
JSPS Fellowships at UTAP and RESCEU, respectively.  This work is
supported in part by Monbusho Grant-in-Aid for JSPS Fellows No.~95209.

\setcounter{section}{0}
\setcounter{equation}{0}
\renewcommand{\thesection}{\Alph{section}}
\renewcommand{\theequation}{\Alph{section}\arabic{equation}}
\section*{Appendices}

\section{Slow-roll conditions}\label{sec:sr}

\noindent
Here we summarize all the conditions assumed in our calculation,
without derivation (One can derive them by differentiating Eq.\ref{2}
several times and using the resultant equations recursively). We
assumed that i) the potential has sufficiently gentle slope, ii) slow
roll has been achieved, and iii) the curvature of $\phi$-space is not
too large and is slowly varying [ i) is a necessary condition for
ii)]. To write the conditions quantitatively, let us define the
``norm''
\beq
\| X^a \| := (X^aX_a)^{1/2}, \quad
\| Y_{ab} \| := (Y_{ab}Y^{ab})^{1/2}
\eeq
of a vector $X^a$ and a symmetric second-rank tensor $Y_{ab}=Y_{ba}$,
and introduce a small quantity $\varepsilon\ll 1$. Then, in the
calculation of $P$, we assumed [conditions i) and iii)]
\bea
&&
\|U_{\!,a}U_{\!,b}\| < \varepsilon,
\quad \| U_{\!;ab}\| < \varepsilon,
\quad \| R_{acbd}U^{\!,c}U^{\!,d} \| < \varepsilon,
\nonumber \\ && \label{ass1}
\| U_{\!;abc}U^{\!,c} \| < \varepsilon^2,
\quad \| (R_{acbd}U^{\!,c}U^{\!,d})_{\!;e}U^{\!,e} \| <
\varepsilon^2
\eea
so that Eq.\ref{38} is valid up to order $\varepsilon$. Also we need
the extra but rather weak condition:
\beq\label{ass2}
\left\| \frac1{H^4}\frac{D^3}{dt^3}\dot\phi^a \right\|
 < \varepsilon^{5/2}
\eeq
because of ii). The second line in Eq.\ref{ass1} is necessary to make
sure that the first order quantities are slowly varying. Similarly, in
order that Eqs.\ref{44} and \ref{45} are valid up to order
$\varepsilon^2$, the calculation of $n$ assumes
\beq\label{ass3}
\| U_{\!;abcd}U^{\!,c}U^{\!,d} \| < \varepsilon^3, 
\quad \| U_{\!;abc}U^{;cd}U_{\!,d} \| < \varepsilon^3, \quad
\| [(R_{acbd}U^{\!,c}U^{\!,d})_{;e}U^{\!,e}]_{;f}U^{\!,f} \|
< \varepsilon^3
\eeq
in addition to Eq.\ref{ass1}, and
\beq\label{ass4}
\left\| \frac1{H^5}\frac{D^4}{dt^4}\dot\phi^a \right\|
 < \varepsilon^{7/2},
\eeq
instead of Eq.\ref{ass2}.


\section{Derivation of Eq.\protect\ref{9}}\label{app:a}
\setcounter{equation}{0}

\noindent
From Eq.\ref{1}, the stress tensor is given by
\beq
T^{\mu}_{\:\nu}= h_{ab}g^{\mu\rho}
\partial_{\rho}\phi^a \partial_{\nu}\phi^b
- \left( \frac12 h_{ab}g^{\rho\sigma}
 \partial_{\rho}\phi^a \partial_{\sigma}\phi^b - V
 \right) g^{\mu}_{\,\nu}\, ,
\eeq
and the Euler-Lagrange equation is
\beq\label{a2}
\left[ \frac{\! D}{dx^{\mu}} + (\partial_{\mu}\ln\!\sqrt{-g}\,) \right]
h_{ab}g^{\mu\nu}\partial_{\nu}\phi^b + V_{,a} = 0
\eeq
which yields Eq.\ref{2}. Choosing the gauge ${\cal R}=B=0$ in
Eq.\ref{11}, the relevant components of the metric perturbation are
(we work in $\vect k$-space)
\beq
\delta g_{00} = -\delta g^{00} = 2A\,, 
\quad \delta g_{0i} = \delta g^{0i}=0\,, \quad
\delta\ln\!\sqrt{-g} = A - k^2 E\,.
\eeq
We perturb $T^{\mu}_{\:\nu}$ covariantly with respect to $\phi^a$ and
define the covariant perturbation operator
$\delta:=\per\phi^a\nabla_{\!\! a}$; for example, $\delta
h_{ab}=0$. Since $\per\phi^a$ is Lie-transported along $\phi^a(t)$,
i.e., $[\per\phi, \dot\phi]^a = 0$, it follows that
\beq\label{a4}
\delta(\dot\phi^a)= \frac{D}{dt}\per\phi^a.
\eeq
Using Eq.D.7 in Ref.\cite{KS} to calculate the perturbation in the
Einstein tensor, $\delta G^0_0 = \delta T^0_{\,0}$ and $\delta
G^0_{\,i}=\delta T^0_{\;i}$ give
\beq
-6H^2 A - 2k^2H \dot E 
= -A\dot\phi^a\dot\phi_a +
 \dot\phi^a \frac{D}{dt}\per\phi_a + \per\phi^a V_{,a}\,,
\eeq
\beq\label{a5}
2HA = \dot\phi_a\per\phi^a,
\eeq
respectively. Using Eq.\ref{2}, one finds
\beq\label{a6}
\dot A + k^2 \dot E = \per\phi^a \frac{D}{dt}
\left(\frac{\dot\phi_a}{H}\right)\,.
\eeq
Perturbing Eq.\ref{a2} covariantly gives
\bea
&&
\left( \frac{D}{dt} + 3H \right) \left(\frac{D}{dt}\per\phi_a 
- 2A\dot\phi_a \right)
 + \left( \delta\frac{D}{dt} - \frac{D}{dt}\delta \right)\dot\phi_a 
\nonumber \\ &&
+ (\dot A - k^2\dot E)\dot\phi_a + \left(\frac{k}a\right)^2\per\phi_a 
+ \per\phi^b V_{;ab} = 0.
\eea
Here the second term is calculated as
\bea
\left(\delta\frac{D}{dt}-\frac{D}{dt}\delta \right)\dot\phi_a 
&=& (\per\phi^b\nabla_{\! b}\dot\phi^c\nabla_{\!\! c}
- \dot\phi^c\nabla_{\!\! c}\per\phi^b\nabla_{\! b})\dot\phi_a 
\\ 
&=& 2 \per\phi^b\dot\phi^c \dot\phi_{a;{\scriptscriptstyle [}c
b{\scriptscriptstyle ]} }
+ \per\phi^b\dot\phi^c{}_{;b}\dot\phi_{a;c}
- \dot\phi^c\per\phi^b{}_{;c}\dot\phi_{a;b}
\\ 
&=& R_{acbd}\dot\phi^c\dot\phi^d\per\phi^b.
\eea
on account of Eq.\ref{a4}. Therefore
\bea\label{a11}
&&
\frac{D^2}{dt^2}\per\phi_a + 3 H \frac{D}{dt}\per\phi_a
+ R_{acbd} \dot\phi^c \dot\phi^d \per\phi^b
+ \left(\frac{k}a\right)^2 \per\phi_a + \per\phi^b V_{;ab} 
\nonumber \\ &&
= (\dot A + k^2 \dot E) \dot\phi_a
+ 2A (\ddot\phi_a + 3H \dot\phi_a )\,.
\eea
From Eqs.\ref{a5} and \ref{a6}, it is easy to show that the RHS of
Eq.\ref{a11} is equivalent to that of Eq.\ref{9}.

\section{Some useful formulae}\label{app:b}
\setcounter{equation}{0}

\noindent
To evaluate Eqs.\ref{37} and \ref{39}, we calculate
\beq\label{b1} 
\frac{d\ln(N^{\!,a}N_{\!,a}H^2)}{d\ln a} = 
2\frac{\dot H}{H^2} + 2\frac{N^{\!,a}\dot N_{\!,a}}{N^{\!,c}N_{\!,c} H}
\eeq
where $\dot N_{\!,a}$ is calculated as
\bea
\dot N_{\!,a} &=& \dot\phi^b \nabla_{\! b}\nabla_{\!\! a} N\\
&=& \nabla_{\!\! a}(\dot\phi^b\nabla_{\! b} N) - (\nabla_{\!\! a}\dot\phi^b)
(\nabla_{\! b} N)\\ 
&=& -H_{,a} -N^{\!,b}\dot\phi_{b;a}\,.\label{b4}
\eea
Taking the gradient of Eqs.\ref{2} and \ref{3}, one obtains
\beq\label{b6}
H_{,a} = \frac1{6H} ( V_{,a} + \dot\phi^b\dot\phi_{b;a})
\eeq
\beq
\dot\phi_{b;a} = -\frac1{3H} [3\dot\phi_b H_{,a} 
+ R_{acbd}\dot\phi^c\dot\phi^d + V_{;ab}
+ \dot\phi^c{}_{;a}\dot\phi_{b;c}+ (\dot\phi_{b;a}\dot) ]\,.
\label{b7}
\eeq
The second term in Eq.\ref{b7} arises when one commutes the covariant
derivatives of $\dot\phi^c\nabla_{\!\! a}\nabla_{\!\!
  c}\dot\phi_b$. From these equations,
$\dot\phi_{a;b}\simeq\dot\phi_{b;a}$ holds to the lowest order and
thus the second term in Eq.\ref{b6} becomes $\ddot\phi_a$. Using
Eq.\ref{2} again, one finds
\beq\label{b8}
H_{,a} \cong -\frac12 \dot\phi_a\,.
\eeq
We use $\cong$ when the equation is valid up to the next
order. Defining the first order quantity
\beq\label{b9}
\gamma_{ab} := \left(\frac12h_{ac}h_{bd} - \frac13 R_{acbd}\right)
\frac{\dot\phi^c\dot\phi^d}{H^2} 
- \frac{V_{;ab}}{3H^2}\,,
\eeq
Eq.\ref{b7} can be written iteratively as
\beq\label{b10}
\dot\phi_{b;a} \cong H\gamma_{ab}
- \frac13 H\gamma_a^c\gamma_{bc} 
-\frac{(H\gamma_{ab}\dot)}{3H}\,.
\eeq
From Eqs.\ref{b1}, \ref{b4}, \ref{b8}, \ref{b10} and \ref{lambda}, one
obtains
\bea
\frac{d\ln(N^{\!,a}N_{\!,a}H^2)}{d\ln a}
&\cong& -2\epsilon_{ab} M^{ab}
+ \frac23 M^{ab}\left(\lambda_a^c\lambda_{bc}
 + \alpha \lambda_{ab} - \frac{\dot\lambda_{ab}}H \right) 
\nonumber \\ &&
- \frac{2\alpha}{N^{\!,c}N_{\!,c}} -\frac43
\frac{N^{\!,a}}{N^{\!,c}N_{\!,c}} 
\frac{\dot\phi^b}H \frac{V_{;ab}}{3H^2}   \label{b11}
\\ &\simeq& -2\epsilon_{ab} M^{ab}\,. \label{b12}
\eea
Also it is easy to
show that
\beq\label{phidot}
\frac{\dot\phi^a}H \cong - (U^{,a} 
+ \frac13 U^{;ab}U_{\!,b}) \simeq -U^{,a}\,,
\eeq
\beq
H^{-1}\dot N_{\!,a} \simeq \lambda_{ab}N^{\!,b}\,,
\eeq
\beq
H^{-1}\dot M_{ab} \simeq 2(
\lambda^c{}_{{\scriptscriptstyle (}a}M_{b{\scriptscriptstyle )}c}
 - M_{ab}M_{cd}\lambda^{cd})\,.
\label{b14}
\eeq


\end{document}